\documentclass[aps,prb,notitlepage,superscriptaddress,reprint,longbibliography]{revtex4-2}
\usepackage{braket}
\usepackage{bm} 
\usepackage[utf8]{inputenc} 
\usepackage{siunitx}
\DeclareSIUnit{\sample}{Sa}
\usepackage{graphicx}
\usepackage{amsmath} 
\usepackage{placeins} 
\usepackage[dvipsnames,svgnames,x11names]{xcolor}
\usepackage{float}
\usepackage{natbib}
\usepackage{textcomp}
\usepackage{siunitx}
\usepackage{scrlayer-scrpage}
\usepackage{svn-multi} 
\usepackage{amssymb}
\usepackage{amsthm} 
\usepackage{xr} 
\usepackage[colorlinks=true,allcolors=Blue]{hyperref}
\hypersetup{pdfprintscaling=None}

	\usepackage[final]{changes}

\newcommand{\coloronline}[1]{{}} 

\begin{document}

\title{Modeling shallow confinement in  tuneable quantum dots}

\author{Austris Akmentinsh}
\affiliation{Department of Physics, University of Latvia, Jelgavas street 3, LV-1004 Riga, Latvia}
\author{Niels Ubbelohde}
\affiliation{Physikalisch-Technische Bundesanstalt, 38116 Braunschweig, Germany}
\author{Vyacheslavs Kashcheyevs}
\email{Corresponding author: slava@latnet.lv}
\affiliation{Department of Physics, University of Latvia, Jelgavas street 3, LV-1004 Riga, Latvia}

\begin{abstract}
This paper proposes a universal microscopic model for the shallow confinement  regime of  single-electron tunneling  devices. We consider particle escape from a  quantum well generically emerging  as a bifurcation in a smooth  electrostatic potential and develop a set of analytic and numerical approximations for the ground-state tunneling and thermally activated escape rates. These approximations are applied to the problem of electron capture by a closing tunnel barrier where the competition between the closing speed and the escape rate defines a scaling relation for the capture fidelity. Effective one-dimensional cubic potential approximation leads to a universal form of this  scaling relation in terms of device-independent dimensionless  depth and speed parameters.  Using predictions for temperature and magnetic-field dependence we show how to infer the energy scales of cubic longitudinal  and quadratic transverse confinement.   
Finally, we derive an intrinsic  quantum speed bound for adiabatic protection of the ground state tunneling and show that the latter can potentially be exploited up to the break down of confinement with a practical speed limit set by reaching the quantum uncertainty of the barrier height before the onset of non-adiabatic excitation.   
These results contribute to mapping out the physical limits of single-electron quantum technologies for electrical metrology and sensing.
\end{abstract}

\maketitle

\section{Introduction and overview}

A confined quantum state isolated from the continuum by a tuneable tunnel barrier is a very generic scenario  in physics~\cite{Ankerhold2007}.  Particular relevance of this problem in the context of electrostatically defined quantum dots (QDs) arises from a steady progress of single-electron quantum technologies where individual electrons, isolated in near-deterministic manner from the bulk Fermi sea, are envisioned to serve as intrinsically calibrated units  for primary electrical metrology or ballistically propagating probes for quantum sensing and quantum information \cite{Pekola2013,Kaestner2015,Bocquillon2014,Bauerle2018,RadarPaper2024}.
Exponential sensitivity of tunneling rates to field effect in semiconductors  is an essential building block of theses technologies as it enables picosecond-scale signal modulation~\cite{Fletcher2012,Yamahata2019PicosecondSource} and high fidelity control of electron number~\cite{Kashcheyevs2010,Reifert2019,Giblin2019}, yet the same sensitivity makes quantitative measurement and modelling of fast tunnel rates challenging.

{A particular difficulty in modeling of electronic  quantum devices~\cite{Edlbauer2022} is countering} the uncertainty  of the electrostatic potential on the microscopic level as the lithographic structures 
and the associated gate potentials vary over length scales larger than the typical size of the localized single-electron wave-function. Mesoscopic unpredictability of frozen disorder in a depleted semiconductor only adds to this challenge \cite{davies1989,yang2009}. Reconstructing the confinement potential directly from quantum transport data may offer a  route for circumventing this problem.

\begin{figure}
    \centering
    \includegraphics[width=6cm]{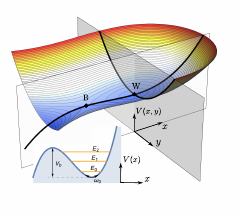}
    \caption{Visualisation of a confining potential of a shallow QD. A minimum (W) defines the bottom of a confining well, and a saddle-point (B) -- the top of a tunnel barrier. The inset shows the cubic approximation to the cut in the longitudinal dimension. $V_b$ is the barrier height and $\omega_0$ is the (cyclic) frequency of linear oscillations. The number of discrete resonances $E_n$ for escape along $x$ is approximately $u=V_b/(\hbar \omega_0)$. 
    \label{fig:fig1}}
\end{figure}

Here we consider one electron in a  QD in which confinement is tuned to be  \emph{shallow}, i.e.,  the curvature of the confining potential  becomes small along one particular dimension (longitudinal direction, labeled $x$) as shown schematically in Fig.~\ref{fig:fig1}.
In this regime a tunnel barrier  (which controls the entrance to the QD) and the potential minimum (which localizes the quantum states inside) are spatially close, and the electrostatic potential can be expanded around the inflection point in the middle suggesting a cubic polynomial in $x$  as a generic approximation. Coefficients of this expansion and the lever arm factors with which external voltages can tune the linear term thus become the key device parameters to be determined.  

In this paper we lay out  a strategy for accessing these microscopic confinement parameters by treating  single-electron tunneling from a cubic  potential as an inverse scattering problem: if the initial condition for a single confined electron can be controlled, accurate measurements of the escape rates $\Gamma$ over a sufficiently wide  range as a function of the external variables (gate voltages, temperature, magnetic field) provide enough information to validate both the cubic approximation and the ground state tunneling assumption, as well as  estimate the parameters.
Application of this strategy has become possible with a recent experimental demonstration \cite{Akmentinsh2023} of universal scaling of integrated escape rates reaching into the shallow regime.
This work complements Ref.~\onlinecite{Akmentinsh2023} with the
 theoretical foundation of the effective 1D cubic potential and detailed discussion of thermal activation. Furthermore, we explore additional  implications of the model,   predicting a scaling of the  effective potential parameters with the magnetic field, and deriving bounds on adiabaticity in the application to the charge capture (single-electron isolation) problem.

The paper is structured into three main parts (Section \ref{sec:microscopic} to \ref{sec:NonAdiabatic}), each addressing different aspects of the problem.
Section \ref{sec:microscopic} is the most general as it seeks to apply the well researched theory of metastability~\cite{Ankerhold2007,Weiss2008,Hanggi1990} to  a cubic potential  describing a QD. We introduce the notation and the characteristic scales for a static cubic potential in $x$ in Sec.~\ref{eq:potentialDefitinition}, discuss the interplay of electrostatic and magnetic confinement in $(x,y)$ plane  and the conditions for reduction to an effective 1D problem in Sec.~\ref{sec:magfield}, and then proceed to describe the discrete decay modes (scattering resonances) of the cubic potential in Sec.~\ref{sec:MicroscopicDecayRates}, paying particular attention to the range of validity of commonly used analytical approximations and the non-perturbative nature of the confinement in a very shallow  (effectively single-level) dot where anharmonicity becomes important.
We conclude the general discussion in Section~\ref{sec:ThermalDecayRates} by considering crossover from tunneling to hopping (thermal activation) within quantum transition state theory approximation, which we adapt to  a few-level regime utilizing the exact results of Sec.~\ref{sec:MicroscopicDecayRates}. As the 1D cubic approximation has been thoroughly tested for \emph{macroscopic} quantum tunneling of the superconducting phase~\cite{Devoret1985,Martinis1987}, in Section~\ref{sec:secSCcomapre} we briefly compare the relevant energy scales between the superconducting phase and the single-electron confinement, and demonstrate the robustness of the particular analytic approximation to thermal activation developed in Sec.~
\ref{sec:ThermalDecayRates}.

The second part of the paper (Section \ref{sec:section3}) introduces dynamics to the switching of tunneling by considering closing an initially shallow QD  with a linear voltage ramp. This represents a well-defined application for the approach developed in Section~\ref{sec:microscopic} and also constitutes a key step in operation of on-demand sources~\cite{Leicht2011,Fletcher2012} of isolated electrons~\cite{Fletcher2019} and metrological charge pumps~\cite{Kaestner2015,Giblin2019}. Results from Section~\ref{sec:microscopic} are used to relate the model parameters to the main measurable quantity --- single-electron capture probability. Scaling of the latter with QD closing speed, temperature and magnetic field is discussed to (a) establish tests for conditions  when electrons predominantly escape by tunneling from the metastable ground state, and (b)  estimate parameters of the  shallow potential, most notably gate voltage dependence of the absolute depth (barrier height) and the energy gap protecting the ground state. In  Section~\ref{sec:magneticTested}, comparison to early literature data~\cite{kaestner2009a,kataoka2011} on non-adiabatic charge pumping in different magnetic fields \cite{Wright2008,Fletcher2012} confirms the promise of the effective 1D potential approximation (Section \ref{sec:magfield}) to make further connections between the microscopic theory and the experimental realizations.

The third part --- Section \ref{sec:NonAdiabatic} --- estimates the limits to quantum adiabaticity intrinsic to the linear ramping scenario considered in Section \ref{sec:section3}. Probabilities for excitation-driven escape are estimated taking into account both squeezing and acceleration of the evolving quantum well and an upper bound on the dimensionless speed parameter for maintaining ground-state tunneling (and hence minimal capture error) is established.
We find that in the shallow limit these theoretical results are consistent with previous investigations of quantum nonadiabaticity in single-level tuneable-tunneling QD models~\cite{Flensberg1999,VKJT2012}. Importantly, the adiabaticity bound is not breached in our experiments~\cite{Akmentinsh2023} which probe the universal speed-depth scaling relation.

We conclude the paper with a summary of potential implications of our work and an outlook for further research and development of single-electron control with tuneable tunnel barriers.

\section{Cubic potential model for a shallow quantum dot\label{sec:microscopic}}

\subsection{Model definition and potential parametrization\label{eq:potentialDefitinition}}

We consider confinement of electrons
in a semiconducting nanostructure
by an externally controlled electrostatic potential which is smooth on the scale of the  wave functions of the confined electrons. Typically, such potentials are created and tuned by voltages on external metallic gates via the field effect. 
For a QD deep enough to confine many modes, it  is natural to model  the confining well (centered at W) and the tunnel barrier (B) connecting QD to an outside reservoir separately. In this deep-dot case, separate quadratic expansions around B and around W would give the corresponding characteristic quantum energy scales (energetic width of the barrier transmission function and the level spacing inside the QD, respectively). In contrast, in the shallow limit both scales emerge together via  
 a bifurcation in the confinement potential from a monotonic  $V(x)$ to one with a maximum at B and  a minimum at W, as depicted in Figure~\ref{fig:fig1}. Here $x$ is the real-space coordinate along the least confined dimension, and $V(x)$ is the electrostatic potential energy of a single electron. In the expansion of $V(x)$  around the bifurcation point the dominating term is cubic, which without loss of generality can be expressed as
\begin{equation}\label{eq:Vbdef}
    V(x)
    =
    b \, x^{3}/3 - F \, x+ V_{b}/2 \, .
\end{equation}
Here $F$ measures the detuning from the   bifurcation (stationary inflection) point, and  forming a QD requires $F>0$. 
The quadratic term is eliminated by measuring $x$ from the inflection point of $V(x)$, and $V_b=4 F^{3/2}/(3 \, b^{1/2})$ is the energy distance between the maximum and the minimum, $V_b= V(-x_0)-V(+x_0)$, with $ x_0=(F/b)^{1/2}$ being half the distance between B and W in Fig.~\ref{fig:fig1}. Smoothness assumption requires $x_0 \ll d$ where $d$ is characteristic distance to the closest gates contributing to $V(x)$.   In our approach, the cubic curvature coefficient $b$ at the bifurcation point ($F=0$) serves as a device-specific parameter. As the external voltages  are detuned from this special point and the potential landscape is deformed, parameters of the best-fit cubic approximation to the real potential will change continuously. Under the separation of scales assumption (barely emerging QD, $x_0 \ll d$) the leading order approximation to this change is to treat $b$ as a  constant and $F$ as a linear function of the external tuning  voltages. The spirit of this crucial approximation is similar to Landau free energy expansion near a critical point, aiming to capture universal aspects of the bifurcation.

For a particle of effective mass $m$, the angular frequency of small oscillations near $x=x_0$ is $\omega_0 = (2/m)^{1/2} \, (b F)^{1/4}$. Harmonicity of the quantum mechanically metastable ground state inside the well is controlled by a dimensionless depth parameter $u = V_b/(\hbar \omega_0)$.
 It is expedient to parametrize the problem in terms of a gate-tuneable (via $F$) dimensionless $u$ and a constant  $\Omega_b=(6 \hbar b^2/m^3)^{1/5}$. All non-trivial scattering properties of the cubic potential (detailed in Section~\ref{sec:MicroscopicDecayRates} below) depend only on the value of $u$; the  constant $\Omega_b$, sets the time and the energy  scales. $\Omega_b$ is the device-specific characteristic confinement scale, determined by the cubic curvature  $b$ of the electrostatic  potential and the effective mass $m$ in the particular semiconductor band.
Indeed, the corresponding single-particle Hamiltonian, $\mathcal{H}=-\hbar^2 \partial_x^2/(2 m)+V(x)$, expressed in terms of a dimensionless $\mathrm{x}=(x-x_0)/l_0$, is
\begin{equation}\label{eq:Hvaiu}
    \mathcal{H}
    =
    \hbar \Omega_b u^{1/5}\left(
        -\frac{1}{2}
        \frac{d^{2}}{d\mathrm{x}^{2}}
        +
        \frac{\mathrm{x}^{2}}{2}
        +
        \frac{\mathrm{x}^{3}}{3\sqrt{6u}}
    \right) \, , 
\end{equation}
where $l_0=\sqrt{\hbar/(m \omega_0)}= x_0 \sqrt{2/(3u)}$ is the harmonic confinement length scale.  The dimensionful  depth $V_b$ and  frequency $\omega_0$  are equal to $V_b=\hbar \Omega_b u^{6/5}$ and $\omega_0 = \Omega_b \, u^{1/5}$, respectively.

We see from \eqref{eq:Hvaiu} that for $u \gg 1$ the harmonic approximation for the lowest states should be adequate while for $u \sim 1$  (the shallow limit) non-perturbative anharmonic effects are expected to be essential as the cubic term is no longer small and the distance scale BW emerging from the bifurcation is comparable to the ground state localization length, $x_0 \sim l_0$.

\subsection{Transverse confinement and magnetic field scaling\label{sec:magfield}}

In addition to pure electrostatic confinement, a magnetic field perpendicular to $x$-$y$ plane can be used to control the relevant electronic modes. 
Description of the emerging quantum dot in terms of the one-dimensional potential $V(x)$ alone requires the motion along the other (transverse) dimensions to remain confined and decoupled from the longitudinal coordinate $x$. Such confinement can be provided either by electric or magnetic field, or a combination of both. Within this subsection we consider the two-dimensional motion explicitly, and look for the conditions for dimensional reduction using    the following Hamiltonian:
\begin{align} \label{eq:2Dexplicity}
    \mathcal{H}_{\text{2D}} & =
    \frac{(p_x-m \, y \,  \omega_c)^2+p_y^2}{2 m}+\frac{m \, \omega_y^2 \, y^2}{2} + V(x) \, . 
\end{align}
Here the strength of magnetic  and transverse electric confinement is parametrized by the cyclotron frequency $\omega_c=e B/m$ and $\omega_y$, respectively,   $(p_x,p_y)=-i \hbar (\partial_x, \partial_y)$ are the canonical momenta operators, the vector potential is chosen along the $x$-axis, and $V(x)$
is given Eq.~\eqref{eq:Vbdef}.   The electrostatic part of $\mathcal{H}_{\text{2D}}$ is illustrated in Fig.~\ref{fig:fig1} as $V(x,y)$.

The vector potential induces a  coupling between $p_x$ and $y$. Nevertheless, as we show below,
the full two-dimensional problem \eqref{eq:2Dexplicity} can be approximated by  the one-dimensional  one \eqref{eq:Hvaiu} with appropriately rescaled effective mass  as long as  $\omega_0 \ll \omega_y$ regardless of magnetic field strength.
This is achieved by a canonical transformation $
\widetilde{\mathcal{H}}=e^{S} \mathcal{H} e^{-S}$ generated by
\begin{align}
    S=\frac{i \, \omega_c \, p_x \, p_y}{m \, \hbar \, (\omega_c^2+\omega_y^2)} \, .
\end{align}

The transformed Hamiltonian is 
\begin{align}
    \label{eq:H2Dtransformed}
    \tilde{H}_{\text{2D}} & = \left( \frac{\omega_y}{\Omega} \right)^2 \frac{p_x^2}{2 \, m} +V\left(x+  \frac{\omega_c \, p_y}{m \, \Omega^2} \right) +  \tilde{\mathcal{H}}_{y} \, ,
\end{align}

where $\widetilde{\mathcal{H}}_{y}=p_y^2/(2 m)+ m \, \Omega^2\, y^2/2$ describes a harmonic oscillator in the transverse direction with a confinement scale $\Omega =\sqrt{\omega_c^2+\omega_{y}^2}$.
We observe that a finite magnetic field renormalizes the effective  mass in the longitudinal direction and mixes in a fraction of the transverse momentum to $x$. 
If the transverse motion is confined to the ground state of $\widetilde{\mathcal{H}}_{y}$ (i.e., the lowest pinch-off mode of the barrier, point B at $x=-x_0$, viewed as a quantum point contact), then the uncertainty of $x$ due to zero point fluctuations of $p_y$ is on the order of $\delta x(B)= \langle  p_y^2 \rangle_y^{1/2} \omega_c/(m \Omega^2)=
(\omega_c/\Omega) \sqrt{\hbar/(2 m \Omega)}$.
This scale must be shorter than the longitudinal harmonic localization length $l_0(B)=\sqrt{ \hbar \omega_y/ (m\, \omega_0 \, \Omega)}$. The condition $\delta x(B) \ll l_0(B)$ is satisfied for any $\omega_c(B)$ if $\omega_0 \ll \omega_y$.

Averaging $\widetilde{H}_{\text{2D}}$ over the ground state of $\widetilde{\mathcal{H}}_{y}$  gives an effective 1D Hamiltonian
\begin{align} 
    \langle \widetilde{H}_{\text{2D}} \rangle_y = & -\frac{\hbar^2}{2  (\Omega/\omega_y)^2 m }\frac{d^2}{dx^2} +b \, x^3/3 \nonumber \\
    & -(F-b \, \, \delta x^2) \, x  + \hbar \Omega/2 \, .
    \label{eq:groundstateprojection}
\end{align}
We see that up to inconsequential constant shifts in  $F$ and the energy reference level, Eq.~\eqref{eq:groundstateprojection} is unitarily equivalent to the one-dimensional Hamiltonian \eqref{eq:Hvaiu}  
with appropriately rescaled paramaters due to $B$-dependent increase in the  effective mass,

\begin{subequations} \label{eq:scaling}
\begin{align}
    \Omega_b(B)& = \zeta^{-6/5}\Omega_b(0)  \, , \label{eq:OemgabofB}\\
    u(B) & =\zeta  u(0) \, ,  \\
    l_0(B) & =\zeta^{-1/2} l_0(0) \, ,
\end{align}
\end{subequations}
 where we expressed a
 dimensionless rescaling factor $\zeta(B)=\Omega/\omega_y=\sqrt{1+B^2/B_0^2}$ in terms of  $B_0=m \omega_y/e$, the crossover field from electrostatic to magnetic confinement.  
In the large magnetic field limit ($\zeta \gg 1$), $x$ {} becomes the guiding centre coordinate of the localized Landau state, and the anisotropy of the shallow QD ensures that the longitudinal confinement scale $l_0(B)$ remains larger than the magnetic length~$\sqrt{\hbar/(m \omega_c)}$.   
Note that oscillations slow down with $B$, $\omega_0(B)=\zeta^{-1} \omega_0(0)$, while the classical depth $V_b$ remains unchanged. 
 
 In the harmonic (large $u$) limit, the dimensional reduction \eqref{eq:groundstateprojection} and the effective mass scaling \eqref{eq:scaling} can be derived from the saddle-potential solution~\cite{Fertig1987} as detailed in Ref.~\onlinecite{Pavlovska2022}, and are consistent with the exact solution of an anisotropic Fock-Darwin model~\cite{Madhav1994,Kyriakidis2005}.

The scaling relations \eqref{eq:scaling} provide a path to quantitatively control cubic longitudinal confinement, estimate the transverse potential $\omega_y$ and test bounds of the simplifying approximations by varying $B$. An example is discussed in Section~\ref{sec:magneticTested} below.

\subsection{Decay rates of metastable eigenstates of a  cubic potential}\label{sec:MicroscopicDecayRates}

The quantum dot described by the cubic potential~\eqref{eq:Vbdef} holds a finite number (on the order of $u$) of discrete metastable quantum states. Here we review their properties beyond the harmonic approximation.
Our goal is to take into account anharmonicity effects as the states inside the QD become progressively less confined when their energies approach the top of the barrier at $E_n \approx V_b $. 
 In this section, we first accurately define the resonant eigenstates and illustrate their physical properties with a non-perturbative numerical calculation following the dilation-transformation  (complex scaling) method described in Ref.~\onlinecite{Yaris1978,Alvarez1988}. These results are compared to the standard Wentzel–Kramers–Brillouin (WKB) approach with an analytically exact action integral to distinguish the breakdown of the approximations  from the physically meaningful effect of  progressively wider resonances merging into a continuous density of states as function of energy.
The net result of this section are the robust numerical results for the resonance energies and widths as function of $u$ as well as two analytic approximations with a quantitative characterization of their validity. This development
 also sets the stage for approximations of the temperature-dependent decay rate that is the subject of the following section (Section~\ref{sec:ThermalDecayRates}).

The cubic potential is characterized~\cite{Caliceti1980,Alvarez1988} by existence of a discrete hierarchy of scattering eigenstates $\psi_n(x)$ that obey the stationary Schrodinger equation with complex energies, $\mathcal{H} \psi_n(x) = (E_n - i \hbar \Gamma_n/2) \psi_n(x)$, decay to $0$ inside the infinite potential wall, $\psi_n(x\to +\infty)\to 0$, and satisfy outgoing-waves-only boundary condition at $x\to -\infty$  (Gamow-Siegert boundary condition~\cite{Siegert1939}). The complex energies of the eigenstates  $\psi_n(x)$ are also the poles of the scattering matrix corresponding to $\mathcal{H}$. As the time evolution of these resonant states corresponds to a spatially uniform exponential decay, $|\psi_n(x,t)|^2=|\psi_n(x)|^2 e^{- \Gamma_n t} $, the set of $\Gamma_n$ can be understood as the spectrum of  escape rates available in dynamics governed by a cubic Hamiltonian.

\begin{figure}
        \includegraphics[width=7cm]{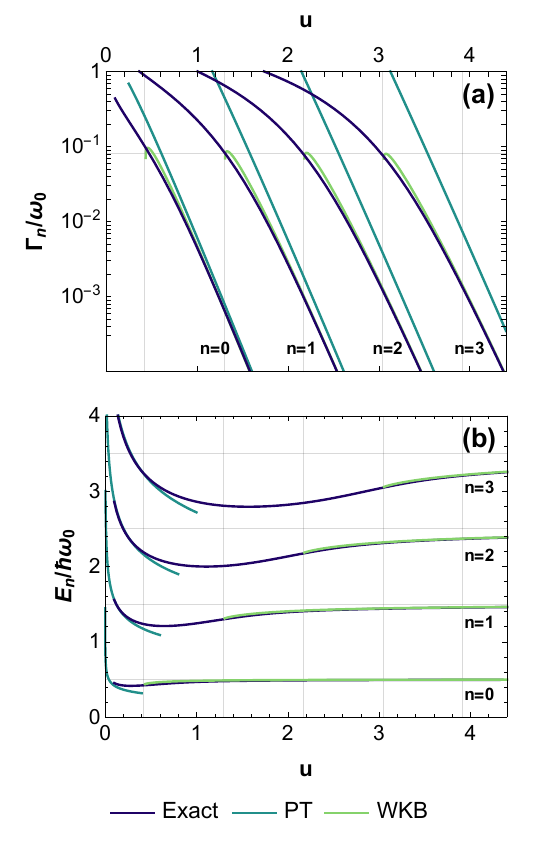}   
        \caption{\label{fig:GammasAndEnergies}
Cubic potential resonance (a) widths (tunneling rates $\Gamma_n$) and (b) positions (energies $E_n$ measured from the classical minimum) computed exactly via complex scaling (dark blue), with the standard WKB method (green) and from the asymptotic perturbation theory (cyan). The slope of $\log \Gamma_n$ converges to $-36/5$ and the energy levels become harmonic at large $u$, in agreement with Eq.~\eqref{eq:PTboth}. Vertical grid lines in both panels mark the values of $u_n=5 \pi (2n+1)/36$ at which another semiclassical bound state, $E_{n}^{\text{WKB}}(u_n)=V_b(u_n)$, becomes available, see Eqs.~\eqref{eq:WKBenergy} and \eqref{eq:botha}.}
\end{figure}

An infinite sequence of resonances exists for any value of dimensionless coupling~\cite{Alvarez1988} which in our parametrization \eqref{eq:Hvaiu} is expressed by $u\ge 0$. The dimensionless widths  $\Gamma_n/\Omega_b$ and energies $E_n/(\hbar \Omega_b)$ enumerated in increasing order by $n=0,1,2 \ldots$ 
 are continuous functions of a single argument $u$ that can be computed numerically using the complex scaling method~\cite{Yaris1978,Alvarez1988}(see the Appendix for implementation details). 
The results are shown in Figure~\ref{fig:GammasAndEnergies}.

Resonances with $n< u$ are localized  inside the dot and their energies follow   
asymptotic perturbation theory (PT)~\cite{CaldeiraLeggett1983,Weiss1983,Alvarez1988} with the leading terms for $n \ll u \to \infty$, 
\begin{subequations} \label{eq:PTboth}
 \begin{align}
    \label{eq:largeuasymp}
    \frac{\Gamma_{n}^{\text{PT}}}{\omega_{0}}
    &=
    \frac{6^{3/2}}{\sqrt{\pi}}
    \frac{432^{n}}{n!}
    u^{n+1/2}
    e^{-36u/5} \, ,  \\
    E_n^{\text{PT}} & = \hbar \omega_0 (n+1/2) \, .
\end{align}
\end{subequations}
The poles in the scattering matrix persist across the transition $n \sim u$ and for $n \gg u$ approach the perturbative asymptotics of the opposite limit~\cite{Alvarez1988},
\begin{align}
\label{eq:smalluasymp}
    E_{n}^{\text{PT}}- i \hbar \Gamma^{\text{PT}}_n/2
    =
{\hbar\Omega_b}    \frac{e^{-i\pi/5}}{2^{4/5}}
    \bigg(
        \frac{5\pi^{3/2}(2n+1)}{3\Gamma(1/3)\Gamma(1/6)}
    \bigg)^{6/5} \, ,
\end{align}
for $u \to 0$.
We note that Eq. \eqref{eq:smalluasymp} is exact for $u=0$ which gives $(\Gamma_0^{\text{PT}})^{-1}= 2.6 \, \Omega_b^{-1}$ the interpretation of the maximal quantum lifetime for a particle localized at the classical stationary point $x=x_0=0$  at $F=0$.

To summarize the standard WKB approximation~\cite{Weiss1983,Ankerhold2007}, we consider  classical confined states  at energies $E$ between $0$ and $V_b$.  For $0< E<V_b$, the equation for turning points of the cubic potential $V(x_i)=E$ yields three solutions 
$x_1 < x_2 < x_0$ and $x_3 > x_0$ , which
mark the boundaries of two classically allowed regions: semi-infinite interval $x <x_1$ and the confined region $x_2 < x < x_3$.
The tunneling rate from a confined state at energy $E$ can be expressed as a product of the classical attempt frequency $1/\tau(E)$ and the transmission probability, expressed as exponential of the bounce action~\cite{Ankerhold2007,Weiss2008} $\widetilde{W}(E)$ for a periodic orbit in the inverted potential, 
    \begin{equation}
        \Gamma^{\text{WKB}}(E)
        =
        \frac{e^{-\widetilde{W}(E)/\hbar}}{\tau(E)} \, .
    \end{equation}
Here $\tau(E)$ is the oscillation period in the well between the  turning points $x_2(E)$ and $x_3(E)$, while $\widetilde{W}(E)=2 \int_{x_{1}}^{x_{2}}\sqrt{2m[V(x)-E]}dx$  is integrated over the classically forbidden region $x_1(E)<x<x_2(E)$.  The period $\tau(E)=dW(E)/dE$ of the real orbit in the non-inverted well can be expressed in terms of regular short action $W(E)= 2 \int_{x_{2}}^{x_{3}}\sqrt{2m[E-V(x)]}\, dx=\oint p \, dx$ which is equal to the phase space area encircled by the periodic motion for $E<V_b$.
We denote this dimensionless area $W(E)/(2 \pi \hbar)=a(E)$.

The inversion symmetry of the cubic potential,
$V(x)=V_b-V(-x)$, leads to the symmetry $\widetilde{W}(E)=W(V_b-E)$ and hence
\begin{align} \label{eq:GammaWKBviaa}
  \Gamma^{\text{WKB}}(E) & =\frac{ \exp\left[-2 \pi a(V_b-E)\right ]} {2 \pi \hbar \, da(E)/dE} \, .
\end{align}

In the semiclassical approximation the discrete spectrum arises from the Bohr-Sommerfeld quantization condition,
\begin{align} \label{eq:WKBenergy}
    a\left(E_n^{\text{WKB}} \right)=n+1/2  \, .
\end{align}
and the standard WKB approximation for $\Gamma_n $
is obtained by computing Eq.~\eqref{eq:GammaWKBviaa} at the roots of Eq.~\eqref{eq:WKBenergy},
$\Gamma_n^{\text{WKB}} =\Gamma^{\text{WKB}}\left(E_n^{\text{WKB}} \right)$. The results are compared to exact complex diagonalization in Figure~\ref{fig:GammasAndEnergies}.

We have used the analytic solution of the cubic equation, $x_i(E)$, and  computer-assisted calculus techniques to identify a compact analytic form for the exact phase space area $a(E)$  in terms of  the hypergeometric function, 
\begin{subequations} \label{eq:botha}
\begin{align}
   & a(E) = \frac{E}{\hbar\omega_{0}} \times
    {}_{2}F_{1}\Big(\frac{1}{6},\frac{5}{6},2,\frac{E}{V_{b}}\Big) \label{eq:aexact} \\
  &  = \begin{cases}
        \frac{E}{\hbar \omega_0} \left ( 1+ \frac{5 \, E}{72 \, V_b} \right ), & \! E \ll V_b , \\
        \frac{u}{2 \pi }
        \left [ \frac{36}{5}+ \frac{E-V_b}{ V_b} \left (1+ \ln \frac{432 V_b}{V_b-E}  \right) \right  ],  & \! V_b-E \ll V_b 
    \end{cases} \label{eq:aexpanded}
\end{align}
\end{subequations}
Using the asymptotic expansion \eqref{eq:aexpanded} for $E \ll V_b$ in Eqs.~\eqref{eq:GammaWKBviaa} and \eqref{eq:WKBenergy} and the Stirling formula for $n \gg 1$ in Eq.~\eqref{eq:largeuasymp} it is straightforward to verify that $\Gamma_n^{\text{WKB}}/\Gamma_n^{\text{PT}} \to 1$ in the limit of $1 \ll n \ll u$ as WKB becomes asymptotically exact in the harmonic limit for large quantum numbers~\cite{Weiss1983}.
The finite discrepancy in the exponential prefactor for the ground state, $\lim_{u \to \infty}\Gamma_0^{\text{WKB}}/\Gamma_0^{\text{PT}}=\sqrt{e/\pi}=0.93$ is well documented for large $u$ (Refs.~\onlinecite{Weiss1983,Ankerhold2007}),
and both  $\Gamma_0^{\text{WKB}}$ and $\Gamma_0^{\text{PT}}$ deviate from the exact $\Gamma_0$  by less than $10\%$ for $u>2.5$. Of course, neither approximation is adequate for small enough $u$: 
$\Gamma_n^{\text{WKB}}$ collapses at $E_n \to V_b$ because the classical round trip time $\tau(E)$ diverges as the turning points $x_1(E)$ and $x_2(E)$ merge at $-x_0$, while the leading order perturbation expansion \eqref{eq:largeuasymp} becomes quantitatively wrong, as can be verified in Figure~\ref{fig:GammasAndEnergies}.

As Fig.~\ref{fig:GammasAndEnergies}(b) shows, quantized energies $E_n^{\text{WKB}}$ computed using the exact phase space area remain remarkably accurate (within $ 6\%$) over the full range of existence of the $n$-th  semiclassical orbit: $a(V_b)=18 \, u/( 5 \pi)$ is finite in contrast to its derivative $\tau(E)$ which diverges logarithmically at $E \to V_b$.  
The exact values of $\Gamma_n(u)$ at $V_b(u)=E_n(u)\approx E_{n}^{\text{WKB}}(u)$ are very close to  $\omega_0(u)/\pi$, consistent with a half-transparent barrier and attempted transmission at frequency of $\omega_0/(2 \pi)$, see the value of $1/(4 \pi)$ marked in Fig.~\ref{fig:GammasAndEnergies}(a).
The subtle nonlinearity of $E_n^{\text{WKB}}(u)$ coupled with exponential sensitivity of $\Gamma^{\text{WKB}}(E)$ to small changes in energy  also explains the good accuracy of $\Gamma_n^{\text{WKB}}(u)$ for $n \geq 1 $ and such $u$ that $1/\tau(u)$ accurately represents the attempt frequency.

We will use the robustness of phase-space area arguments for low $n$ and $u$  to define a tailoring procedure between the continuum and the discrete spectrum for the temperature-dependent metastability in the following section.

\subsection{Thermal activation of particle escape in the  shallow confinement limit} \label{sec:ThermalDecayRates}

One of the ways to probe the spectrum of escape rates is to consider thermally activated escape. 
In this section we adopt to the shallow limit the analytical approximations of the quantum transition state theory \cite{Miller1975,Hanggi1990}, following the main lines of the semiclassical approach of Affleck~\cite{Affleck1981}  (see also Weiss and Haeffner~\cite{Weiss1983} and reviews in Refs.~\onlinecite{Hanggi1990,Ankerhold2007}).
In this approach to thermal activation the coupling to a heat bath is not considered explicitly but instead quantified by a single temperature $T=\beta^{-1}/k_B$. The meaning of this $T$ for a single electron is to assign Boltzmann weights $e^{-\beta E}$ to states of different energy $E$ in a statistical average for a prediction of the escape rate:
\begin{align}
    \langle \Gamma \rangle =Z^{-1} \int \Gamma(E)  \rho (E) e^{-\beta E} \, dE  \, , \label{eq:basicQTST}
\end{align}
where $\rho(E)$ is the density of states in the QD region, $\Gamma(E)$ is the energy dependent escape rate, and $Z=\int \rho(E) e^{-\beta E} dE$ is the partition function.

\begin{figure}
    \includegraphics[width=7cm]{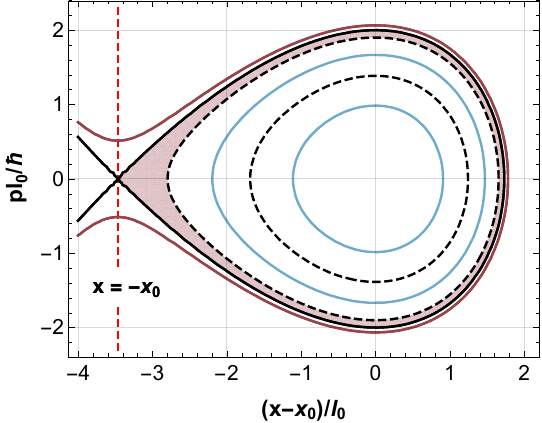}
    \caption{\label{fig:QDInPhaseSpace}
    Phase space for motion in a cubic potential \eqref{eq:Vbdef} of depth  $u=2$.  Constant total energy $E(x,p)$ lines (classical trajectories) are shown for  the resonant energies $E=E_n$ (closed trajectories in blue for $n=0,1$ and an open trajectory in red for $n=2$). The separatrix at $E=V_b$ (black continuous line)  encircles the domain of closed trajectories to the right of the potential maximum (at $x=- x_0$, the red dashed line). Trajectories encircling integer area are marked by dashed lines; the remaining shaded region  has an area  $a_{2}=0.292$, defining the fractional  spectral weight of the last discrete resonance at $n=n_b=2$ in the thermal average in Eqs.~\eqref{eq:DiscreteSums}. }.
\end{figure}

We split the thermally averaged escape rate into two parts:
\begin{equation}\label{eq:ThermalGammaSplit}
    \langle\Gamma\rangle
    =
    \langle\Gamma\rangle_\text{discr}
    +
    \langle\Gamma\rangle_\text{cont} \, ,
\end{equation}
where
$\langle\Gamma\rangle_\text{discr}$ is the contribution from the discrete states of energy $E_n$ localized in the QD and $\langle\Gamma\rangle_\text{cont}$ is the contribution from the continuum of states with energies $E>V_b$ above the barrier. The partition function is split accordingly, $Z=Z_\text{discr}
    +
    Z_\text{cont}$. These two contributions correspond to partitioning the QD phase space (half-plane at $x>-x_0$) into two complementary domains of closed and open trajectories, respectively,  as illustrated in Fig.~\ref{fig:QDInPhaseSpace}.

The discrete part of the thermal average is computed as
\begin{subequations} \label{eq:DiscreteSums}
\begin{align}
    \langle\Gamma\rangle_\text{discr}
    & =
    Z^{-1}
    \sum_{n=0}^{n_b}
            a_n e^{-\beta E_{n}}\Gamma_{n}
\label{eq:GammaDiscr}    \\
    Z_\text{discr}
    & =
    \sum_{n=0}^{n_b}
            a_n e^{-\beta E_{n}},
            \label{eq:Zdiscr}
\end{align}
\end{subequations}
where $\Gamma_{n}$ and $E_{n}$ are the decay rates and energies discussed in Section~\ref{sec:MicroscopicDecayRates}, and the weights are $a_{n<n_b}=1$ and $0 \leq a_{n_b}  < 1$. The number of fully contributing  resonances, $n_b= \lfloor a(V_b) \rfloor =\lfloor 18 u /( 5 \pi) \rfloor $, is computed as the number of full phase space area quanta inside the classically confined region [cf.~Eq.~\eqref{eq:WKBenergy} and Fig.~\ref{fig:QDInPhaseSpace}]. The fractional weight of the last contribution to the discrete sums \eqref{eq:DiscreteSums} is  given by the residual area $a_{n_b} = a(V_b)-n_b$, see the shaded region in Fig.~\ref{fig:QDInPhaseSpace}.
As the semiclassical energy quantization, i.e.~Eq.~\eqref{eq:WKBenergy}, is pretty accurate, we have $a_{n_b}\approx 0.5$ for such $u$ that the topmost contributing resonance is aligned with the top of the barrier,  $E_{n_b}=V_b$.

The contribution of the continuum of unbounded trajectories at $E >V_b$ is computed in a standard way~\cite{Affleck1981} from Eq.~\eqref{eq:basicQTST}, taking into account $2 \pi \hbar \rho(E) \, \Gamma(E) =  \mathcal{T}(E)$,
\begin{subequations}
\begin{align}
    \langle\Gamma\rangle_\text{cont}
    & =
    \int_{V_b}^{\infty}
        \mathcal{T}(E)
        e^{-\beta E}
    \frac{dE}{2\pi\hbar Z}
    \label{eq:ThermalGammaCont}
    \\
    \mathcal{T}(E)
    & =
    \frac{1}
    {1+\text{exp}[-2\pi(E-V_{b})/\hbar\omega_{0}]}
\end{align}
\end{subequations} 
where $\mathcal{T}(E)$ is the barrier transmission probability, taken here in the quadratic approximation near the top of the barrier. For the continuum contribution $Z_{\text{cont}}$ to the partition function $Z$ we evaluate the phase-space  integral over the whole QD area, $x >-x_0$, and then subtract the contribution of closed trajectories (already accounted for in 
$Z_{\text{discr}}$) inside the domain of finite motion, $E(x,p)=p^2/(2 m) +V(x)<V_b$. This recipe gives 
\begin{multline}\label{eq:ContPartitionFunction}
    Z_\text{cont}
 =
    \frac{1}{\sqrt{2 \pi\beta \hbar \omega_0 }}
    \int_{-\sqrt{6 u}}^{\infty}
        e^{-\beta \hbar \omega_0 \left [\mathrm{x}^2/2
        + \mathrm{x}^3/ (3 \sqrt{6 u}) \right ]  }
    d\mathrm{x} \\
    -
    \int_{0}^{V_b}
        {}_{2}F_{1}\left(
        \frac{1}{6},\frac{5}{6},1,\frac{E}{V_{b}}
    \right) e^{-\beta E}  \frac{dE}{\hbar\omega_0} \, , 
\end{multline}
where  
the semiclassical density of confined states per unit energy, $\rho(E) =da/dE$, in the  range of their existence, $0<E<V_b$, 
is expressed analytically using Eq.~\eqref{eq:aexact}.

Tayloring of the two energy ranges above and below the barrier 
via the 
factional weight $a_{n_b}(u)$ 
and the use of numerically exact $\Gamma_n$ and $E_n$  up to $n=n_b$ constitutes our adaptation of Eq.~\eqref{eq:basicQTST} to the shallow limit. This ensures an accurate approximation to the escape rate $\langle \Gamma \rangle$ across  the crossover~\cite{Affleck1981,Weiss1983} from tunneling at $T<T_0=\hbar \omega_0/(2 \pi k_B) $ to thermally activated hopping at $T>T_0$, with the low- and high-temperature limits of $\Gamma_0$ and $\omega_0 \, e^{-\beta V_b}/(2 \pi) $, respectively.
Within the crossover  region $T \approx T_0 \pm \delta T $ of characteristic width~\cite{Affleck1981} $\delta T/T_0 \approx 0.15/\sqrt{u}$ the dominating contribution to the statistical average \eqref{eq:GammaDiscr} shifts from the lowest energies near the ground state  to the topmost states  near $E \approx V_b$.

\begin{figure}
    \centering
 \includegraphics[width=7cm]{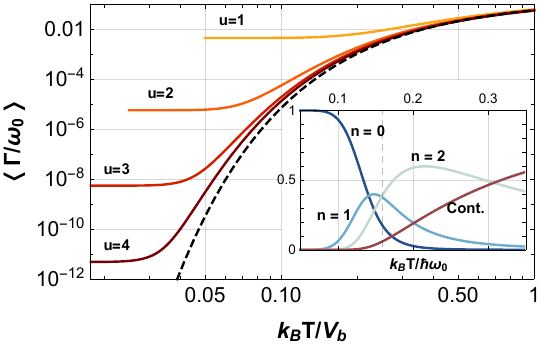}
\caption{\label{fig:isothermal} Temperature dependence of the escape rate $\langle \Gamma \rangle $ according to quantum transition state theory adapted to shallow cubic confinement, Eq.~\eqref{eq:ThermalGammaSplit}. Individual  colored curves correspond to a set of depth parameters $u=1,2,3,4$, the dashed line --- to the classical limit (Arrhenius law) $\langle \Gamma \rangle = \omega_0 \, e^{-\beta V_b}/(2\pi)$. Saturation to the ground-state tunneling rates is clearly observed below a crossover. The inset:
relative  weights of the discrete state  ($n=0,1,2$) and the continuum contributions in the overall thermal average of $\langle \Gamma \rangle$ (normalized to sum to 1). The depth is $u=2.61$ corresponding to $n_b=2$ and $a_{0,1,2}=1.0$; the vertical dashed line marks the crossover temperature $k_B T_0/(\hbar \omega_0)=1/(2\pi)$.}
\end{figure}

This behaviour is illustrated in Fig.~\ref{fig:isothermal} where $\langle\Gamma\rangle$ is plotted as function of temperature for a number of fixed values of the depth parameter $u$.  Our approximation for the cubic potential accurately captures both the low and the high-temperature limits and also illustrates the widening of the crossover region between the two as $u$ is lowered. 
In the inset
we show the fractional weights of the four different contributions to the thermal average for $u=2.61$: three discrete resonances, $\Gamma_n e^{-\beta E_n}/(Z \langle\Gamma\rangle)$ for $n=0,1,n_b=2$, and the continuum contribution, $\langle\Gamma\rangle_{\text{cont}}/\langle\Gamma\rangle$.
We note that above the crossover temperature the last term in the discrete sum \eqref{eq:GammaDiscr} (the topmost resonance) becomes the largest  as the thermal activation to the top of the barrier dominates over ground-state  tunneling.

\subsection{Relation to macroscopic quantum tunneling\label{sec:secSCcomapre}}

The focus of our work are implications of the cubic potential \eqref{eq:Hvaiu} to the physics of shallow quantum dots induced in a depleted semiconductor conduction band. Yet another particular instructive example  to which most of the discussion in the present Section equally applies is tunneling of  the superconducting phase in circuit quantum electrodynamics~\cite{Blais2021}. For a current biased Josephson junction, the  bifurcation point of the effective potential for the phase is known as the critical point of the junction at which the phase becomes deconfined and a voltage drop develops.

An  ideal Josephson junction with a critical current $I_c$ and capacitance $C$, biased by a constant current $I$,  close to the transition  to the finite voltage state, $I \to I_c$, is described by a cubic potential~\cite{Fulton1974,Martinis1987} with the barrier height $V_b=(4 \sqrt{2}/3) E_J \, (1-I/I_c)^{3/2}$  and the (cyclic) plasma frequency $\hbar \omega_0=2^{7/4}\sqrt{ E_J \, E_C}(1-I/I_c)^{1/4}  $ where $E_J=\hbar I_c/(2 e)$ and $E_C=e^2/(2 C)$ are the Josephson and the single-electron charging energies, respectively~\cite{Krantz2019}. Hence
\begin{align}
    u & = (2^{3/4}/3) \sqrt{E_J/E_C} (1-I/I_c)^{5/4}\label{eq:uInJosephsonJunction}
    \, , \\
    \hbar \Omega_b & = 
2^{8/5} (3 E_C^3 E_J^2)^{1/5} \approx 3.78 \, E_C^{3/5} \, E_J^{2/5} \, . \label{eq:OmegabInJosephsonJunction}
\end{align}

The cubic  approximation of the tilted washboard potential requires $1-I/I_c \ll 1$, therefore accessing the shallow confinement regime $u\sim 1$ of the superconducting phase requires a junction with low zero-point-fluctuations of the phase, $\sqrt{E_C/E_J} \ll 1$. For a modern transmon qubit~\cite{Krantz2019} with $E_C/(2 \pi \hbar)=(100 - 300) \SI{}{MHz}$ and $\sqrt{8 E_C \, E_J}/(2 \pi \hbar) =(3 - 6)\SI{}{GHz} $
one gets $\hbar \Omega_b = 10 - \SI{20}{\mu eV} $.

We observe that ramping of the bias current $I$ in the Josephson junction implementation is equivalent to ramping a gate voltage controlling the depth of the QD, as
both $F$ and $I_c-I$ are proportional to $u^{4/5}$.

\begin{figure}
    \includegraphics[width=7cm]{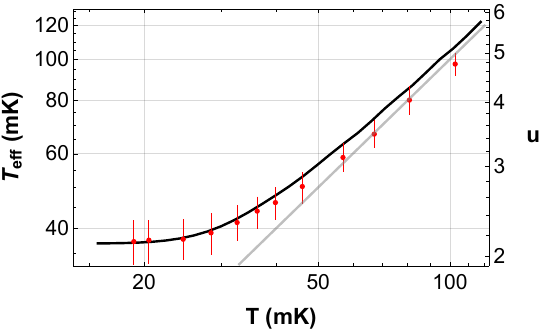}
    \caption{Comparison of experimental data on superconducting phase tunneling in a low-dissipation Josephson junction from Ref.~\onlinecite{Devoret1985} (dots) with our model for thermally activated escape, Eq.~\eqref{eq:ThermalGammaSplit}(black line). 
        The effective temperature $T_{\text{eff}}(T)$ is defined by postulating $\langle \Gamma \rangle= \omega_0 e^{- V_b/(k_B T_{\text{eff}})} / (2\pi)$ for any bath temperature $T$. For each experimental point, $V_b$ and $\omega_0$ are adjusted by tuning the bias current $I$ according to  Eq.~\eqref{eq:uInJosephsonJunction} while keeping 
        $\text{ln}\big( 2\pi\langle \Gamma \rangle / \omega_0 \big) = 11$ which implies $k_B T_\text{eff} = V_b(T)/11$.
         Straight gray line corresponds the pure Arrhenius behaviour $T_\text{eff} = T$ and the right vertical axis indicates the values of $u$ corresponding to $V_b(T)$. \label{eq:figureMartinis}
    }
\end{figure}

Pioneering experiments on macroscopic quantum tunneling~\cite{Devoret1985,Martinis1987,Kagan1992} have worked with $\omega_0/(2 \pi)$ accessible to microwave spectroscopy. In particular, the experiments~\cite{Devoret1985} demonstrating  saturation of phase jump rates (and hence, macroscopic quantum tunneling) at tens of mK temperature  have employed a low-dissipation junction for which the quantum transition state theory is adequate~\cite{Kagan1992,Ankerhold2007}. Using our calculation of $\langle \Gamma \rangle$ described in Section~\ref{sec:ThermalDecayRates} to fit the data for the ``quantum junction'' (quality factor $Q=30$)  from Figure 2 of Ref.~\onlinecite{Devoret1985}, we find good consistency at parameter values
$\hbar \Omega_b=\SI{14}{\mu eV}$ and $u=2.1$ for the ground-state tunneling. These data  also allow to illustrate the crossover to thermally activated escape, as shown in Fig.~\ref{eq:figureMartinis}.

\section{Probing  shallow confinement regime in single-electron tunneling devices\label{sec:section3}}

\subsection{Charge capture by a linear voltage ramp\label{sec:CaptureByLinearRamp}}

Driving the confinement potential across the bifurcation point $F=0$ with a gate voltage ramp is  a generic scenario for formation of a dynamic quantum dot that can trap a discrete number of elementary charges. This approach to single-electron control ensures high accuracy of the number of isolated electrons~\cite{Kashcheyevs2010}, and has been very fruitful for development of high-accuracy current standards \cite{Pekola2013,Kaestner2015} and on-demand electron sources \cite{Bauerle2018,Edlbauer2022}. 

A particularly robust regime in which the applicability of the cubic potential model for shallow confinement can be studied  quantitatively~\cite{Akmentinsh2023} is the escape of the last electron  triggered by a sharp lifting of the Pauli blockade and a gradual closing of the dot with a linear gate voltage ramp.  Such an idealized capture protocol can be summarized as follows:
\begin{enumerate}
    \item At the initial time $t=0$ the dot contains  a single electron ($N_0=1$) with other electrons expelled by a large addition energy ($E_c > V_b$).
    \item For $t>0$, the initially confined electron is able to tunnel out of the QD while the depth-controlling linear term in the confining potential \eqref{eq:Vbdef} increases linearly with time, $F=F_0+ \dot{F} \, t$. In terms of $u$, this defines an initial depth $u_0$ corresponding to $F_0$ and a dimensionless speed parameter $\dot{u}/\omega_0=5  \dot{F}  m/(6 b \hbar)$ which is time-independent as both $\dot{u}$ and $\omega_0$ are proportional to $F$ to the same power, $\dot{u} \propto \omega_0 \propto F^{1/4} \propto u^{1/5}$.
    In terms of an ac gate voltage implementing the ramp, $V_\text{S}(t)$, and a dc gate voltage tuning the starting depth, $V_\text{D}$, relation to the dimensioless model parameters requires device-specific lever arm factors, $\dot{u}/\omega_0 \propto \dot{V}_{\text{S}}$ and $F_0 \propto (u_0)^{4/5}=\tilde{\alpha} (V_\text{D}-V_\text{D}^{\text{c}})$. 
    
    \item The average number of electrons $\langle N \rangle$  to stay on the dot at $t\to \infty$ is measured by repeating the capture protocol many times, and inferring the trapped charge by a separate circuit (either with dc current measurements or electron counting). If by choosing $F_0$ the number $N$ is confined to be either $0$ or $1$, then $\langle N \rangle$ is a direct measure of the single-electron capture probability.
\end{enumerate}
A distinct feature of our modelling approach is identification of the linear expansion term ($-F x$ in the cubic potential Eq.~\eqref{eq:Vbdef}) as the one which is controlled by gate voltages with constant lever arm factors.

Neglecting re-population possibility during escape, the
capture probability follows from the definition of the escape rate,
\begin{align} \label{eq:captureDirect}
    \langle N \rangle = N_0 \, \exp \left [- \int_0^{\infty} \Gamma(t) d t \right ] \, ,
\end{align}
with $N_0=1$ established by an earlier loading stage~\cite{Kaestner2015,Yamahata2021,Hohls2022}. The main condition for a well-defined initial moment (step 1 above) is a strong plunger function of the gate that creates the tunnel barrier~\cite{Yamahata2014,VKJT2012}, i.e.\ fast rising of the overall energy level [zero reference of Eq.~\eqref{eq:Vbdef}] above the Fermi level of the source lead which triggers the backtunneling  (step 2) on a timescale sharper than the subsequent evolution of $\Gamma(t)$.

\subsection{Universal scaling for the ground state tunneling
\label{sec:scaling}}
Taking one electron suddenly out of equilibrium with the source makes the capture process non-adiabatic with respect to particle number equilibrium (this is the defining principle of non-adiabatic charge pumping~\cite{moskalets2002B,Kaestner2007c,Kaestner2015}).
As for the subsequent evolution of the internal electronic state of the QD, it may or may not adiabatically follow the external driving, depending on the speed with which the QD is closed off. In our model, this speed is controlled by the parameter $\dot{u}/\omega_0$.
In the adiabatic limit $\Gamma(t)$ is determined by the instantaneous value of $F(t)$ with no memory or speed-dependent excitation. In this case the speed and the initial depth dependencies of the integrated escape rate in Eq.~\eqref{eq:captureDirect} factorize and the capture probability at different speeds should follow a single scaling function,

\begin{align} \label{eq:Madiabatic}
    M(u_0) = \frac{\dot{u}}{\omega_0} (-\ln \langle N \rangle ) = \int_{u_0}^{\infty} 
\frac{\Gamma(u)}{\omega_0(u)}  \, du  \, ,
\end{align}
which is obtained from Eq.~\eqref{eq:captureDirect} by changing the integration variable from $t$ to $u$ and moving the constant $\dot{u}/\omega_0$ to the front of the integral.
In the tunneling limit escape is dominated by the ground state tunneling rate $\Gamma_0$ (see Section~\ref{sec:MicroscopicDecayRates} above) and adiabaticity is protected by the intradot excitation gap (see Section~\ref{sec:NonAdiabatic} below). The corresponding universal function $M(u)=M_0(u)$ is obtained by substituting $\Gamma(u)$ in Eq.~\eqref{eq:Madiabatic} with $\Gamma_0(u)$ and integrating numerically.

Away from the strong unharmonicity regime, 
the asymptotic form of $M_0(u)$ can be obtained  from Eq.~\eqref{eq:largeuasymp},
$M_0(u_0) \sim 5 \sqrt{u_0/(6 \pi)}\, e^{-36 \, u_0/5}$ which is equivalent to 
 $M_0 \approx  5 \Gamma_0/(36 \omega_0)$. For large $u_0$ we also have $d(u_0^{4/5})/du_0 \approx \text{const}$ within an experimentally accessible range, thus in this regime the ground state capture probability can be  directly approximated by a double exponential as a function of the tuning gate voltage $V_{\text{D}}$,
 $\langle N \rangle=\exp[-\exp(-\alpha V_{\text{D}} +\delta_1)]$  with $\delta_1 = -\ln (\dot{u}/\omega_0) +\text{const}$. In this way we recover the WKB-inspired ansatz postulating $M \propto \exp[-\alpha V_{\text{D}}]$ directly (known in the applications to quantized charge pumps as part of the decay cascade model~\cite{Kashcheyevs2010,Kaestner2015,Giblin2019}).

In  the accompanying paper~\cite{Akmentinsh2023} we describe an experimental protocol to verify validity of adiabatic scaling \eqref{eq:Madiabatic} over many orders of magnitude and deduce the empirical scaling curve. We then fit the latter to $M_0(u_0)$  to deduce the level arm factor $\tilde{\alpha}$, the bifurcation  gate voltage $V_{\text{D}}^{\text{c}}$, and the absolute scale for the dimensionless speed parameter $\dot{u}/\omega_0$ for a given device.
Crucial for this deduction is the non-linearity of $\ln M_0$ as function of $V_D$ which comes from both the slight nonlinearity of $\ln M_0(u_0)$ versus $u_0$ and the non-integer power law $u_0^{4/5} \propto V_D$, most pronounced at $u_0 \lesssim1$. This feature distinguishes  predictions of the inflection point linearization approach from the WKB-inspired approximation $\ln M_0 \propto V_D$ of  the decay cascade model where scaling of the ramp speed and a shift in the control voltage $V_D$ are equivalent~\cite{Fujiwara2008}.

\subsection{Temperature dependence of capture probability}

In Section~\ref{sec:ThermalDecayRates} we have considered thermal activation for particle escape from a static shallow potential. The  Boltzmann averaging  approach, Eq.~\eqref{eq:basicQTST}, reduces the complexity of interaction between the captured electron and a fluctuating external bath to a single extra parameter, $kT/(\hbar \Omega_b)$. This implicitly assumes   a thermalization timescale, controlling  how quickly an equilibrium distribution is established between the energy levels available to the one particle.
In the static case, time evolution of this distribution is irrelevant: after an electron in a particular energy state escapes  into an empty lead, a fresh initial  distribution is prepared in the next cycle.

In contrast, for a time-dependent problem  redistribution between energy levels competes with the rate of change of the confinement potential itself. Staying on the level of simple approximations, we consider two opposite limits. In the fast thermalization limit, the mixing over the internal states is faster than the change in the escape rates, $\langle N \rangle_{\text{fast}} = \exp [- \int \langle \Gamma \rangle ]$ where $\langle \Gamma \rangle$  is a quasistatic equilibrium average following the slowly changing instantaneous depth. The opposite limit is that of slow thermalization,  $\langle
N \rangle_{\text{slow}}= \langle \exp[ -\int \Gamma ] \rangle$ where the 
averaging is done against the initial state ensemble, disregarding  population redistribution between the evolving intradot energy levels. From thermodynamic perspective, 
in the fast thermalization limit we assume good thermal contact (isothermal process), in the opposite limit -- no heat exchange (isoentropic process).

For the specific capture protocol described in Section~\ref{sec:CaptureByLinearRamp}, the two opposite approximations for thermally activated escape are constructed as follows. In the fast thermalization limit,  the rate $\langle \Gamma \rangle$ is given by  Eqs.~\eqref{eq:ThermalGammaSplit} and \eqref{eq:Madiabatic} with time-dependent potential parameters. As this rate is completely determined by the instantaneous value of $u$ and the temperature $T$, the adiabatic scaling relation  \eqref{eq:Madiabatic} holds, so that $\langle N \rangle_{\text{fast}} =\exp[-(\omega_0/\dot{u}) M_T]$ where the temperature-dependent scaling curve  $M=M_T (u_0, kT/\hbar \Omega_b)$ is obtained by replacing $\Gamma(u)$ in Eq.~\eqref{eq:Madiabatic} with $\langle \Gamma \rangle$.

For the slow thermalization limit of the capture  protocol,
consider an initial Boltzmann distribution, $p_n = e^{-E_n(u_0)/kT}/Z$ over a finite number of levels, $n=0, 1, \ldots n_b =\lfloor 18 u_0 /( 5 \pi) \rfloor$, set by the initial depth $u_0$. The partition function $Z$, also computed at $u=u_0$, is the sum of  the discrete \eqref{eq:Zdiscr} and the continuum  \eqref{eq:ContPartitionFunction} contributions. The capture probability is then computed as 
\begin{align}
    \langle N \rangle_{\text{slow}} = \sum_{n=0}^{n_b} a_n \, p_n e^{-\int_{0}^{\infty} \Gamma_n dt }= \sum_{n=0}^{n_b} a_n  p_n 
    e^{-(\omega_0/\dot{u}) M_n(u_0)} \, . \label{eq:slow}
\end{align}
where $M_n(u_0)$ denotes the dimensionless integral on the r.h.s.\ of Eq.~\eqref{eq:Madiabatic} for the specific excited state rate $\Gamma_n$. The fractional weight $a_{n_b}(u_0)$ of the topmost resonance makes $\langle N \rangle_{\text{slow}}$ continuous with respect to $u_0$; inclusion of $Z_{\text{cont}}$ in the normalization of $p_n$'s means that the fraction initially excited into the continuum is assumed to always escape.

\begin{figure*}
    \centering
    \includegraphics[width=15cm]{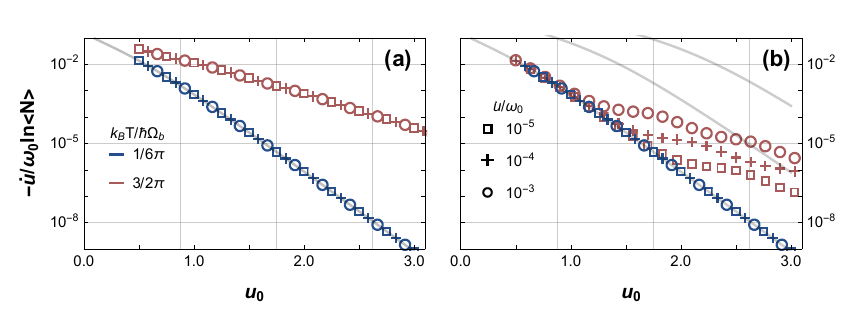}
    \caption{\label{fig:MasterCurveExistence} Charge capture probability $\langle N \rangle$ at different temperatures (color-coded two values of  $k T/(\hbar \Omega_b)$) and capture speeds (shape-coded three values of $\dot{u}/\omega_0$). (a) Fast thermalization model, defined by integrated Boltzmann-averaged escape rate, obeys the adiabatic scaling \eqref{eq:Madiabatic} with a speed-independent $M_T(u_0, kT/\hbar \Omega_b)$. (b) Slow thermalization approximation, Eq.~\eqref{eq:slow}, which fully neglects internal relaxation and  activation during escape, fails to scale with the speed the way capture with adiabatic rates does, cf.~panel (a). Continuous lines show scaling curves corresponding to a particular resonance: (a) ground state $M_0(u_0)$, and (b) ground and excited states $M_n(u_0)$ with $n=0,1,2$.
    }
\end{figure*}

The effect of temperature on the charge capture probability in the two opposite limits is shown in Figure~\ref{fig:MasterCurveExistence}. We plot both  $\langle N \rangle_{\text{fast}}$ and $\langle N \rangle_{\text{slow}}$ in a way that would reveal a speed-independent scaling function $M$; as expected, only the fast thermalization limit in panel (a) does. In both limits, there is an activation threshold for the slope of $u_0$-dependence of $\ln(-\ln \langle N \rangle)$: for $k_B T < 0.16 \hbar \Omega_b$ the capture probability follows the zero-temperature scaling curve $M_0(u_0)$. This can be used to estimate $\Omega_b$  from measurements~\cite{Akmentinsh2023}.

We note that the manifestation of the thermal activation effect discussed here is consistent with a recent theoretical study by Yamahata \emph{et al.}~\cite{Yamahata2021} where charge capture by a multi-level finite-depth QD has been explored employing fast-thermalization and harmonic confinement approximations. In comparable regimes, both studies attribute a crossover temperature in the slope of $\ln (-\ln \langle N \rangle$) (for $\langle N \rangle \to 1$, i.e. large $u_0$) to a  QD-intrinsic energy scale, which we associate quantitatively with the cubic bifurcation scale $\Omega_b$.

There is a depth- and speed-dependent non-equilibrium effect in 
Figure~\ref{fig:MasterCurveExistence}(b): once $T$ is above the activation threshold, 
 different-temperature $\langle
N \rangle_{\text{slow}}$ converge to the ground-state scaling curve at shallow depths (small $u_0$). This difference of $\langle N \rangle_{\text{slow}}$ from the thermally well-coupled limit $\langle N \rangle_{\text{fast}}$ can be attributed to depletion of the initially Boltzmannian population of excited states due to faster escape. This effect is similar to energy-diffusion-limited starvation of the population near the  top of the barrier in the low-dissipation limit of Kramers' transition rate  theory~\cite{Hanggi1990}. We also plot in Figure~\ref{fig:MasterCurveExistence} hypothetical scaling curves $M_n(u_0)$ that would correspond to a deterministically prepared $n$-th excited state [putting $\Gamma_n$ in Eq.~\eqref{eq:Madiabatic}]. Contrasting these with the full Boltzmann-weighted average \eqref{eq:slow} (the symbols in  Figure~\ref{fig:MasterCurveExistence}) gives a visual estimate of the relevance of the thermally excited decay channels (cf.\ inset of Fig.~\ref{fig:isothermal}).

The two approximations to thermal activation, contrasted in Fig.~\ref{fig:MasterCurveExistence}, have the advantage of simplicity of a single parameter $kT/(\hbar \Omega_b)$ characterizing the degree of thermal excitation.  Yet one should keep in mind the limitations in applying either of the two to model quantitatively the temperature dependence in real experiments. 
An explicit model of the electron-phonon coupling would be necessary to describe the competition between the fluctuation-dissipation strength, the rate of escape and the rate of driving. Such coupling at a moderate level is implicitly assumed in the transition state theory underlying the quasistatic Boltzmann distribution in our fast thermalization model,  $\langle N \rangle_{\text{fast}}$.  In terms of the QD physics, this assumes the phonon-driven relaxation rate to be faster than the relevant $\Gamma_n$ yet not too strong for the level broadening to destroy the discrete resonances.

\subsection{Scaling of capture probability with the magnetic field\label{sec:magneticTested}}

Using the results of Sec.~\ref{sec:magfield} and Eq.~\eqref{eq:Madiabatic} for the ground-state tunneling gives
$\langle N \rangle = \exp [- \zeta^{-2} M_0(\zeta u_0) \, \omega_0/\dot{u}  ]$ where
the initial depth $u_0$ and the speed parameter $\dot{u}/\omega_0$ refer to $B=0$ values and the magnetic field dependence is determined by $\zeta(B) = \sqrt{1+B^2/B_0^2} \geq 1$.  
This result implies sharpening of the transition
from $\langle N \rangle =0$ to $\langle N \rangle =1$ as function of the depth-controlling voltage $V_D$ with increasing magnetic field.
Such effect has been observed~\cite{Wright2008,kaestner2009a,Wright2012} and exploited~\cite{Giblin2012,Kaestner2015} in metrological current sources operating in the backtunneling-dominated regime where the current $I = e f \langle N \rangle$ is produced by repeating the capture process at a suffciently high frequency $f$.

Fletcher \emph{et al.}~\cite{Wright2012} have attributed qualitatively the increase in charge capture accuracy with the magnetic field  to the scaling of the effective barrier thickness in proportion to the ground state confinement length of an isotropic Fock-Darwin model.
Our derivation gives this sharpening effect a more explicit microscopic interpretation: according to Eq.~\eqref{eq:H2Dtransformed}, increasing $B$ increases the effective mass, hence reduces the level spacing and increases the sensitivity of the initial number of confined levels $u_0$ to the magnetic-field-independent initial barrier height (set by $V_{D}$).  

The scaling described in Sec.~\eqref{sec:magfield} can be tested  against the available data on single-electron pumps\cite{kaestner2009a,kataoka2011} as follows. For a well-localized ground state we have 
$M_0 \approx  5 \Gamma_0/(36 \omega_0)$ (better than 10\% accuracy for $u_0>1.2$), and taking derivative of $\langle N \rangle$ at the position of the step gives 

\begin{align} \label{eq:derivative}
      \left . \frac{ d \langle N \rangle}{d V_D} \right \rvert_{\langle N \rangle=0.5} 
      &  =\left . \frac{ d \langle N \rangle}{d u_0} 
      \right \rvert_{\langle N \rangle=0.5} 
      \frac{d u_0}{d V_D}    
   \propto  \zeta(B)  \, . 
\end{align}
In the last step  we can disregard the weak nonlinearity of $u_0(V_D)$ dependence as $u_0 \gtrsim 1$ and 
the range of $u_0$ corresponding to current level of $0.5 ef$ that will be probed at different $B$ and fixed pumping frequency $f$ is limited.

\begin{figure}
    \centering
    \includegraphics[width=7cm]{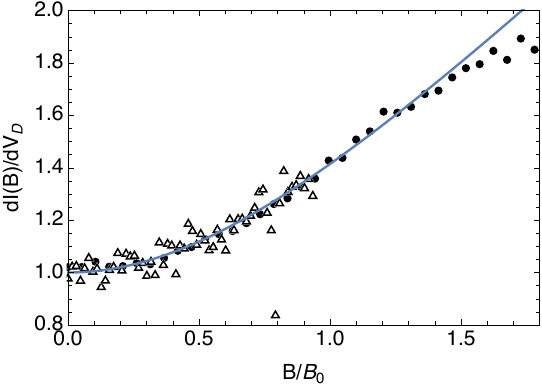}
    \caption{\label{fig:Magnetic}
    Slope  $dI/dV_D$ of the single-electron backtunneling-limited current quantization steps  at $I=0.5 ef$ from two experiments (datapoints)  shown as the function of magnetic field $B$ and normalized by  the value at $B=0$,
    compared to $\sqrt{1+B^2/B_0^2}$  expected from \eqref{eq:derivative}. Data from Ref.~\onlinecite{kaestner2009a} Figure 1(b), $f=\SI{50}{MHz}$ up to $B=\SI{3}{T}$ (triangles) and Ref.~\onlinecite{kataoka2011}  Figure 2(a),  $f=\SI{100}{MHz}$ up to $B=\SI{3.5}{T}$, scaled with $B_0= \SI{3.1}{T}$ and  
   $B_0=\SI{1.9}{T}$, respectively.}
\end{figure}
In Figure \eqref{fig:Magnetic} we compare Eq.~\eqref{eq:derivative} to two datasets from early investigations of GaAs non-adiabatic charge pumps. The inferred $B_0$ of $2$ to $ \SI{3}{T}$ implies transverse confinement 
scale of $\hbar \omega_y = 3-\SI{5}{meV}$ for the particular device designs. 

The scaling \eqref{eq:derivative} predicted from the ground state tunneling evolution will be eventually limited by the non-adiabatic excitation effects as the adiabaticity-protecting scale $\Omega_b$ goes asymptotically to zero with large $B$, see Eq.~\eqref{eq:OemgabofB}.

\section{Adiabaticity conditions for tunnel barrier closing\label{sec:NonAdiabatic}}

Quantum adiabatic theorem~\cite{Born1928} protects the ground state of a slowly evolving quantum system against excitation. For the tunneling out of the metastable ground state of the  cubic potential, the protected state is the lowest resonance wavefunction, $\psi_0^{\text{ad}}(x,t)$, considered in the frozen-time (infinitely slow or adiabatic) limit. As long as the actual wavefunction remains close to $\psi_0^{\text{ad}}$, the time-dependent escape rate $\Gamma(t)$ also stays close to instantaneous $\Gamma_0$ dictated parametrically by the shape of the potential [i.e., by the instantaneous value of $F(t)$].  This is the quantum meaning of the adiabaticity condition underlying the scaling relation \eqref{eq:Madiabatic} for time-integrated tunneling rates  as explained in Section~\ref{sec:scaling}. The energy gap $E_g(t)$ protecting the adiabatic quantum state evolution is the excitation energy to higher resonances, $E_g \sim \hbar \omega_0$ or, in the shallow limit ($V_b < \hbar \omega_0$), to the continuum above the barrier $ E_g \sim V_b$. 

In this section we consider the linear driving protocol for closing the QD modelled by the the 1D cubic potential \eqref{eq:Vbdef} with a linearly increasing force term $F(t) \propto t$,  discussed extensively in Section~\ref{sec:CaptureByLinearRamp}. As explained above, for the charge capture fidelity $\langle N \rangle$ to be  limited by ground-state backtunneling, the electron must remain close to the parametrically evolving instantaneous ground state of the QD, $\psi_0^{\text{ad}}(x,t)$. 
Here we first solve the internal quantum dynamics of the weakly excited confined state in the harmonic approximation and derive \emph{a priori} upper bounds on the speed parameter $\dot{u}/\omega_0$ for the  adiabatic scaling relation 
\eqref{eq:Madiabatic} for the ground state rate, $M_0(u)$. The method relies on the exact solution for quantum dynamics of a driven harmonic oscillator~\cite{Husimi1953} and can also be adapted to other driving protocols. At the end of the section, we treat the shallow limit separately, using a Landau-Zenner-type argument of Kashcheyevs and Timoshenko~\cite{VKJT2012}, and arrive at a consistent upper bound $\dot{u}/\omega_0 <0.1$ for all relevant $u$.

We start the analysis from the 
Newton's equation of motion $m \ddot{x} =-\partial V(x,t)/\partial x$ which in the reference frame following the minimum $x_0(t)$ of the cubic potential~\eqref{eq:Vbdef}, is expressed via $\xi(t) = x(t)-x_0(t)$ as
\begin{align}\label{eq:Newton}
    m \ddot{\xi} =-m \ddot{x}_0(t) - m \omega_0^2(t) \xi(t) - b\, \xi(t)^2
\end{align}
with $\omega_0^2 = 2\sqrt{b F(t)}/m \propto t^{1/2}$ and $x_0 =\sqrt{F(t)/b} \propto t^{1/2}$  (see also Sec.~\ref{eq:potentialDefitinition}). In this section we reference $t$ from the moment of bifurcation, $F(0)=0$.
Staying at rest at the bottom of the well, 
$\xi^{\text{ad}}(t)\equiv 0$, is not an exact solution to \eqref{eq:Newton} because of the non-zero inertial force  $m \ddot{x}_0(t)$. 
Rather, a non-oscillating solution to \eqref{eq:Newton} with the smallest amplitude,
$\xi^{\text{min}}(t)$, asymptotically approaches the adiabatic classical equilibrium 
at large $t$ as $\xi^{\text{min}}(t) \sim m /( b t^2) \to \xi^{\text{ad}}(t)=0$. In the harmonic approximation [i.e., neglecting the $\xi^2$ term in Eq.~\eqref{eq:Newton}], this minimally excited classical solution can be expressed analytically as
\begin{align} \label{eq:ximin}
    \xi^{\text{min}}(t)/ x_0(t) & =  \frac{\pi}{5} \sqrt{2-2 / \sqrt{5}}
    \left[ J_{-2/5}(z)+ J_{2/5}(z) \right ] \nonumber \\
   & - {}_1F_2\left(1 ; \frac{4}{5} ,  \frac{6}{5}  ; -\frac{z^2}{4} \right ) 
   \equiv \Xi(z)  \, ,
\end{align}
where $z\equiv \omega_0 \, u/\dot{u} =4 \, \omega_0(t) \, t/5$, $J_{\nu}(z)$ is the Bessel function of the first kind and  ${}_1F_2$ is the generalized hypergeometric function. The latter cancels the oscillatory behavior of the particular combination of Bessel functions in Eq.~\eqref{eq:ximin}. 
Asymptotically,  $\Xi(z) \sim 4/(25 z^2)$  at large $z$.

As $2 |x_0(t)|$ is the distance between the top of the barrier and the bottom of the well and $\xi(t)$ is the displacement from the bottom, the harmonic approximation is accurate for $\xi^{\text{min}}(t)/ x_0(t) \ll 1$ and remains qualitatively valid up to  $\xi^{\text{min}} (t)/ x_0(t)  \sim 1 \to z \gtrsim 0.06$ or $\dot{u}/\omega_0 < 17 \, u$. Below we analyze internal quantum non-adiabaticity in the harmonic approximation.

As first observed by Husimi~\cite{Husimi1953}, an exact Gaussian wave-function for a time-dependent harmonic oscillator takes the form
\begin{align} 
    \psi(\xi,t) =& \exp \left[\frac{i m \dot{\xi}_2(t)}{2 \hbar \xi_2(t)} \left \{ \xi-\xi_1(t) \right \}^2 \right ] \nonumber \\
   &  \times \exp \left [i m \dot{\xi}_1(t) \left \{ \xi- \xi_1(t)\right \}/\hbar+C(t) \right ]  \label{eq:Husimi} \, ,
\end{align}
where $\xi_2(t)$ is a particular classical solution of a force-free   oscillator, $\ddot{\xi}_2+\omega_0(t)^2 \, \xi_2=0$, and $\xi_1(t)$ is a solution to Eq.\eqref{eq:Newton} in the harmonic approximation but including the non-homogeneous driving (inertial force) term; $C(t)$ is fixed by  normalization and the absolute phase (i.e., time-dependent energy) reference.

Contrasting Eq.~\eqref{eq:Husimi} with the adiabatic ground state wavefunction (in co-moving coordinates $\xi$),
$ \psi_0^{\text{ad}}(\xi,t) =\exp \left [ -m \omega_0(t) \xi^2/(2 \hbar) \right ] $, one can distinguish two kinds of non-adiabatic excitation:
(a) time-dependent displacement of a coherent state in  phase space  from the ground state $\{ 0, 0 \}$ to  $\{ \xi_1(t), m\dot{\xi}_1(t) \} $ in one-to-one correspondence with the classical trajectory $\xi_1(t)$; (b) 
squeezing that is driven by time-dependent confinement strength $\omega_0(t)$ and is quantified by deviations of $\xi_2(t)$ from a fixed-frequency harmonic motion.

We estimate the minimal amount of non-adiabatic excitation by computing the overlap between the Gaussian \eqref{eq:Husimi} with minimal-excitation initial condition [such that the final condition
$\psi(\xi,t) \to \psi_0^{\text{ad}}(\xi,t)$ is satisfied asymptotically] 
and the first and the second excited adiabatic states,  $\psi_{1,2}^{\text{ad}}(\xi,t)$.
This means choosing the initial displacement and squeezing  via the initial conditions for $\xi_1(t)$ and $\xi_2(t)$ such that
reversible time evolution quenches all excitation
at large times. These fine-tuned initial conditions do not represent actual dynamics yet serve as a way to estimate the minimal amount of excitation inherent to the bifurcation scenario for the emergence of the QD expressed by our core assumptions of the cubic potential, $V(x,0) \propto x^3$, and the linear force, $F\propto t$. We note that additional accelerations and squeezing not captured by this scenario may be present, hence the following estimates should be viewed as an upper bound on the speed of the QD formation and closing.

The solution for $\xi_1(t)$ describing the classical particle asymptotically at rest is $\xi_1(t)=\xi^{\text{min}}(t)$; it has already been introduced above in Eq.~\eqref{eq:ximin}. 
The particular solution for $\xi_2(t)$ ensuring that the Gaussian width parameter  $\dot{\xi}_2/\xi_2$ converges to the adiabatic limit of 
$i \omega_0(t)$ at large $t$ is given by 
$\xi_2(t)= (\omega_0 t)^{2/5} H^{(1)}_{2/5}(4 \omega_0 t/5)$. Here $H_\nu^{(1)}(z)$ is the Hankel function of the first kind, a complex linear combination of real-valued Bessel functions $J_{\nu}$ and $Y_{\nu}$, analogous to the particular solution $e^{i z}$ for a unit frequency  harmonic oscillator. Using these particular functions $\xi_1$ and $\xi_2$, the overlap between the Gaussian $\psi(\xi,t)$ given by \eqref{eq:Husimi} and the first and the second excited adiabatic states
can be computed straightforwardly. The analytic result in the small excitation limit,  $p_n^{\text{min}} \ll 1$, is
\begin{gather} 
    p_1^{\text{min}}  = \lvert \braket{\psi_1^{\text{ad}} | \psi^{\text{min}} } \rvert ^2 \approx \frac{m}{2 \hbar} \left (\xi_1^2 \omega_0 +\dot{\xi}_1^2/\omega_0 \right ) \nonumber \\
     =\frac{3 u}{4} \left [ \left (  1+\frac{4}{25 \, z^2} \right) \Xi^2+ \frac{4\, \Xi}{5 z} \frac{d\Xi}{dz}  +  \left (   \frac{d\Xi}{dz} \right )^2 \right ] \, , \label{eq:p1min} 
    \end{gather}
    \begin{align}
    p_2^{\text{min}} &=  \lvert \braket{\psi_2^{\text{ad}} | \psi^{\text{min}} } \rvert ^2 \nonumber \\ & \approx \frac{1}{8} \left \lvert 1+ i\frac{H^{(1)}_{-3/5}(z)}{H^{(1)}_{2/5}(z)} \right\rvert ^2 \, .
    \label{eq:p2min}
\end{align}
At large $z$, $p_1^{\text{min}} \sim 12 \, u/(625 z^4)$ and $p_2^{\text{min}} \sim 1/( 800 z^2)$.
As one would  expect to the lowest order, displacement affects the first excited state most [Eq.~\eqref{eq:p1min}] while squeezing is an even (in $\xi$) perturbation and first becomes pronounced in $p_2$, 
Eq.~\eqref{eq:p2min}.

The effect of excitation on the electron escape problem is to open additional escape channels, in this case -- higher metastable states. 
Similar to crossover to thermal hopping (discussed in Sec.\ref{sec:ThermalDecayRates}), we estimate the threshold for the non-adiabatically driven escape by considering the conditions 
\begin{align} \label{eq:pGamma}
    p_n^{\text{min}} \, \Gamma_n < \Gamma_0 \, , \; (n=1,2) \, .
\end{align}
As $\Gamma_n/\omega_0$ is a function of $u$ only and $p_{1,2}^{\text{min}}$ given by Eqs.~\eqref{eq:p1min}-\eqref{eq:p2min} are functions of $u$ and $z=u/ (\dot{u}/\omega_0)$, we can estimate the upper bounds on the dimensionless speed parameter $\dot{u}/\omega_0$ by solving \eqref{eq:pGamma} for $\dot{u}/\omega_0$  at fixed $u$.
At sufficiently large $u$, the leading order perturbation theory expressions for $\Gamma_{1,2}^{\text{PT}}$ can be used, see Eq.~\eqref{eq:largeuasymp},
which results in the conditions
\begin{align} 
\dot{u}/\omega_0  & < 5 \sqrt{u}/(6 \sqrt{2}) \approx 0.6 \sqrt{u} \,  & & u>1.2 \\
    \dot{u}/\omega_0 & < 5/54 \approx 0.09 & & u>2.7 \label{eq:bound2}
\end{align}
The conditions on $u$ for the above speed limits have been estimated by requiring that the limits computed  with exact $\Gamma_n$ and $p_n^{\text{min}}$ are 
at most twice the stated perturbative values; this is also consistent with requiring a sufficient depth for a well-localized state with $n=1$ and $n=2$, respectively.
We see that at large $u$ (for QD with at least three quantum levels), squeezing is more detrimental to adiabaticity than inertial displacement (i.e., the bound from $p_2^{\text{min}}$ is tighter), unless additional accelerations are present.

When the QD is at its shallowest,  $u \sim 1$, quantum excitation proceeds directly into the continuum. This non-adiabatic excitation mechanism has been considered in single-level quantum dot models for charge capture before~\cite{Flensberg1999,VKJT2012}. We adopt a simple two-level Landau-Zenner estimate~\cite{VKJT2012} for non-adiabatic transition probability $p_c$ over an energy gap $E_{g}$ due to exponentially decreasing tunnel coupling, $p_{\text{c}}=\exp({-\pi \tau E_g/\hbar})$,
where $\tau = -\Gamma(t)/\dot{\Gamma}(t)$ is characteristic decoupling time. In our problem, the gap to continuum is the barrier height, $E_g\approx V_b=u \hbar \omega_0$, and $\tau \approx 5/(36 \dot{u})$ from $\Gamma_0^{\text{PT}}(u)$. A speed limit is reached once this excitation probability exceeds the probability of direct tunneling, $\approx \Gamma_0 \tau$. Solving $p_c < \Gamma_0^{\text{PT}} \tau$ for $\dot{u}/\omega_0$ yields
\begin{align} \label{eq:tocont}
    \dot{u}/\omega_0 < \frac{5 u \pi}{36 W_0(\pi u \omega_0/\Gamma_0^{\text{PT}})} \, , 
\end{align}
where $W_0$ is Lambert $W$ function's principal branch. With Eq.~\eqref{eq:largeuasymp}, the r.h.s.\ of Eq.~\eqref{eq:tocont} is $0.125$, $0.09$ and $25 \pi/1296 \approx 0.06$ for $u=0.5, 1$ and $\infty$, respectively. This  estimate is consistent with Eq.~\eqref{eq:bound2}. 

We conclude that an upper bound on the modulation speed parameter for ground-state tunneling adiabaticity  is approximately $\dot{u}/\omega_0 < 0.1$.
Our estimation for transition threshold to photo-excited escape is very similar to crossover temperature considerations (see Section~\ref{sec:ThermalDecayRates}), and reaffirms~\cite{Flensberg1999} quantitative interpretation~\cite{VKJT2012} of the inverse decoupling time $k T_{\text{eff}}=\hbar /{(\pi \tau)}$ as an effective temperature.

The intrinsic upper bound on the speed parameter $\dot{u}/\omega_0<0.1$ can be combined with  the necessary condition for capture $u_0>0$ to mark the boundaries in the parameter space where the adiabatic scaling relation for the ground state, Eq.~\eqref{eq:Madiabatic} with $M_0(u_0)$,  can be expected to hold. Owing to generic nature of the cubic approximation in the shallow regime, mapping  QD implementations on such universal speed-depth diagram may prove useful in optimizing devices towards the intrinsic quantum speed limit analyzed in this Section. 

\section{Summary and outlook}

We have considered cubic confinement for the last electron in an electrostatically created shallow QD, focusing on the ground-state tunneling regime and ways to connect microscopic parameters with external control variables. Together with the experimental techniques and the data confirming the universal regime of the model~\cite{Akmentinsh2023} this opens new paths for exploration of controlled single-electron tunneling technologies.

A natural next step is first-principles modelling of two-electron interaction in a shallow quantum dot. Extensive literature on  deep quantum dots~\cite{Reimann2002} provides an array of methods and benchmarks to adapt to this situation. Effective shallow potential for the second-last electron and the confinement of the last electron would give a quantitative theoretical prediction for maximal accuracy of tunneling-based electron number quantization~\cite{Kashcheyevs2010,Giblin2019}.  Connecting these limits to the basic device parameters (effective mass, dielectric constant and lithographic geometry) may enable theory-driven optimization and comparative analysis of different  platforms for single-electron manipulation.

Another application of the cubic potential model for shallow confinement is the  on-demand electron emission problem~\cite{Feve2007,Leicht2011,Fletcher2012,Fletcher2019}. The interplay of energy-time uncertainty and adiabaticity of emission has been the topic of a variety of theoretical investigations (time-dependent tunneling Hamiltonian models~\cite{Keeling2008,Kashcheyevs2017,Ubbelohde2015,Contreras2012,Splettstoesser2024}, rate equations~\cite{Leicht2011,Fletcher2019}, numerical solution of time-dependent wave-packet dynamics \cite{Ryu2016,Splettstoesser2024}). Having a microscopic potential model backed by measurements on the capture side may allow more realistic predictions of the shape of emitted wave-packets and potentially enable 
generation of picosecond scale single-electron wavepacket for quantum sensing~\cite{RadarPaper2024,Fletcher2022,Ubbelohde2022,Wang2022}.

\begin{acknowledgments}
   We thank Akira Fujiwara, Frank Hohls,  Nathan Johnson and Peter Silvestrov for discussions, and  Bernd Kaestner and Masaya Kataoka for access to the experimental data of Refs.~\onlinecite{kaestner2009a} and \onlinecite{kataoka2011}.
   A.A. and V.K are supported by grant no.~lzp-2021/1-0232 from the Latvian Council of Science~and the Latvian Quantum Initiative within European Union Recovery and Resilience Facility project no.~2.3.1.1.i.0/1/22/I/CFLA/001.
    This work has been supoprted in part by project 23FUN05 AQuanTEC  which has received funding from the European Partnership on Metrology, co-financed from the European Union’s Horizon Europe Research and Innovation Programme and by the Participating States
\end{acknowledgments}

\section*{Appendix}

Here we summarize the method of complex scaling   \cite{Alvarez1988,Yaris1978} (also known as complex dilation) which we use to compute resonance energies $E_n$ and tunneling rates $\Gamma_n$ of the cubic potential presented in Fig.~\ref{fig:GammasAndEnergies} (for $n$ up to $3$) and available  for download~(up to $n=7$)\cite{ResonanceData2024}.

The basis for the method is the group of 
scaling transformations, $\psi(x)\to e^{a/2} 
 \psi(e^a\,x)${} 
  with a pure imaginary scale parameter $a=i \theta$. For the position $x$ and the momentum $p_x=-i \hbar \partial_x$ operators this implies $x\to e^{i \theta} x$ and $p_x \to e^{-i \theta} p_x$, making the transformed Hamiltonian $\mathcal{H}_{\theta}$ non-hermitian. The key property of the method~\cite{Alvarez1988,Yaris1978} is that for certain intervals of $\theta$ the  outgoing resonance eigenfunctions $\psi_n(x)$ of $\mathcal{H}$ transform into normalizable eigenfunctions $\psi_n(e^{i \theta} x)$ of $\mathcal{H}_{\theta}$,  with the  $\theta$-independent eigenvalues $E_n-i \hbar \Gamma_n /2$.
Thus ideally one only has to find the spectrum of $\mathcal{H}_\theta$ and identify its $\theta$-independent eigenvalues. 
 
 In a practical numerical calculation, we do this by expressing $\mathcal{H}_\theta$ in a finite basis of Hermite functions (harmonic oscillator eigenstates) localized at the bottom of the well, and then  diagonalize the resulting matrix. Due to the finite basis approximation, the eigenvalues $(E_n - i \hbar \Gamma_n/2) / \hbar\omega_0$ acquire a small dependence on $\theta$ and one has to look for the value at which they change the least. One has to control the convergence of these approximated numerical eigenvalues with respect to the basis size and the numerical precision. Our results~\cite{ResonanceData2024}  have been computed with the standard double precision and the basis size of 200. For appropriately selected set of $u$'s we have repeated the calculation with variable numerical precision up to 48 digits and the basis size of $300$. In this way we estimate a relative error of $10^{-4}$ or less for all the $\Gamma_n$ included in the dataset. For $u \to 0$ the numerical spectrum converges to the exact values given by Eq.~$\eqref{eq:smalluasymp}$.


%

\end{document}